\documentclass{pasj00}

\begin{document}
\SetRunningHead{R.~Yamazaki et al.}
{Giant flare of SGR 1806-20 from a relativistic jet}
\Received{2005/03/dd}
\Accepted{2005/mm/dd}

\title{Giant Flare of SGR 1806-20 from a Relativistic Jet}

 \author{%
   Ryo \textsc{Yamazaki}\altaffilmark{1},
   Kunihito \textsc{Ioka}\altaffilmark{2},
   Fumio \textsc{Takahara}\altaffilmark{1},
   and
   Noriaki \textsc{Shibazaki}\altaffilmark{3}}
 \altaffiltext{1}{Department of Earth and Space Science, Osaka University, 
                  Osaka 560-0043, Japan}
 \email{ryo@vega.ess.sci.osaka-u.ac.jp, takahara@vega.ess.sci.osaka-u.ac.jp}
 \altaffiltext{2}{Physics Department and Center for Gravitational Wave 
                  Physics, Pennsylvania State University, PA 16802, USA}
 \email{kunihito@gravity.psu.edu}
 \altaffiltext{3}{Department of Physics, Rikkyo University,  
                  Tokyo 171-8501, Japan}
 \email{shibazak@rikkyo.ac.jp}


\maketitle

\begin{abstract}
Japanese magnetospheric explorer GEOTAIL recorded a detailed light 
curve during the initial 600 msec of a giant 
flare from SGR~1806-20 on December 27, 2004.
We show that the observed 
light curve is well explained by an emission from relativistically 
expanding fireballs, like those of gamma-ray bursts (GRBs). 
Especially, the observed rapid 
fading after 500~msec suggests that ejecta is collimated in a jet. 
We derive an upper limit on the jet opening half-angle of 0.2 radian 
that is as narrow as those of GRBs.

\end{abstract}

\section{Introduction}

Soft gamma-ray repeaters (SGRs)  are most likely highly magnetized 
neutron stars, so called magnetars \citep{thompson95}. 
On December 27, 2004, a giant flare from SGR~1806-20 illuminated the 
Earth \citep{boggs04,hurley05,palmer05} with the gamma-ray flux
more than $\sim 10^{6}$ times that of the
 typical cosmological gamma-ray bursts (GRBs) that are the 
most violent explosions in the universe. 
The giant flare has an initial spike lasting about 600~msec with an
isotropic equivalent energy of $\sim10^{46-47}$~ergs, that is followed by
a pulsating tail lasting 400~seconds with energy of $\sim10^{44}$~ergs.
Only two giant flares from SGRs had been recorded before
this flare: they occurred on 
March~5, 1979 from SGR~0526-66 \citep{cline80} and
August~27, 1998 from SGR~1900-14 \citep{hurley99}, respectively.
The initial spike of the most recent flare is $\sim 10^2$ times more 
energetic than the previous two events,
while  pulsating tails of the three events have comparable energies
(e.g., \cite{woods04}).

Because of  the observed high flux density, most $\gamma$-ray detectors were 
saturated except for particle detectors such as
GEOTAIL that successfully recorded a burst light curve in the 
brightest initial spike
(\cite{terasawa05}; see also \cite{mazets05}).
The burst was so bright that the light curve was clearly recorded down 
to three orders of magnitude below the peak flux. In the early epoch 
($t<160$~msec), the light curve is variable and mainly consists of two pulses. 
A gradual power-law like decay begins after the second peak ($t>160$~msec) 
with a bump at $t\sim 430$~msec. This is followed by a rapid fading after 
$t\sim 500$~msec. Such a detailed light curve of the initial spike has 
been measured for the first time.
It brings us a new key to 
understanding of the mysterious, poorly understood SGRs.

The detection of radio afterglows after giant flares suggests that SGRs 
 eject relativistic outflows \citep{cameron05,gaensler05,frail99}. 
A relativistic motion is also implied by the nonthermal flare spectrum
observed by \citet{mazets05} and \citet{palmer05},
otherwise  pair formation occurs in a compact emission region,
which makes thermal spectrum \citep{huang98,thompson01,nakar05}.
Even if the spectrum is thermal \citep{hurley05},
its hyper-Eddington luminosity implies  a relativistic motion \citep{wang05}.
The initial spike and the pulsating tail have different spectral features 
and temporal pulse profiles, which suggests that they have different
origins: the initial spike and the radio afterglow may arise from the
relativistic outflow, and the pulsating tail
 may come from evaporating trapped fireballs.
The kinetic energy of the outflow inferred from the radio afterglow
is $\sim10^{44-45}$~ergs \citep{cameron05,gaensler05}, which
is much smaller than the isotropic equivalent energy of the initial
$\gamma$-ray spike.
Hence a collimated relativistic outflow is implied.
In addition the isotropic equivalent energy of the initial spike is comparable
to that of the exterior magnetic field  $B\sim10^{15}$~G of the magnetar.
Then the magnetar cannot produce giant flares repeatedly 
(e.g., $\sim 100$ times)
during its active time ($\sim10^4$~yrs) unless
the emission is collimated or there is other energy reservoir.
Therefore we have fair motivations to consider an anisotropic giant flare.

If the giant flares of SGRs arise from relativistic collimated outflows,
they are similar to canonical GRBs
from the cosmological distance (typically tens of Gpc). A sub-group of GRBs 
with long duration is thought to be caused by relativistic jets that originate 
in the collapse of a massive star \citep{hjorth03,stanek03}. Energy is 
carried away from a compact source as kinetic energy of jets
\citep{piran99,piran04,zhang04}.  This is 
converted into radiation by internal shocks between shells,
which make observed highly variable 
gamma-ray light curves, called prompt emissions of GRBs. 
Subsequently, at larger radii, 
outflow interacts with ambient circumstellar matter, producing external 
shocks, which are responsible for afterglows on much longer time scales in 
various wave lengths, such as radio, optical, and X-ray bands.

In this Letter, we show that the observed light curve of the initial
600~msec is well explained by the emission from relativistically 
expanding fireballs, like those of GRBs. Especially, the observed rapid 
fading after 500~msec  suggests that ejecta is collimated in a jet. 
We derive a robust upper limit on the jet opening half-angle of 0.2 radian 
that is as narrow as those of GRBs.

\section{Light curve of collimated outflow}

The light curve of the initial spike of the 
giant flare from SGR~1806-20 is very similar 
to the behavior in prompt GRB emissions (see Figure~\ref{fig:lc}). 
We can interpret two pulses in 
the early epoch ($t < 160$ msec) as two internal shocks (see below). The 
following decay is basically determined by the relativistic kinematics, 
which is independent of the emission mechanism. Suppose a relativistic 
shell shines for a short period. Since the shell has a curvature, photons 
far from the line of sight (LOS) come later. The shell at higher latitude 
from the LOS has a lower velocity toward the observer, so that the emission 
becomes dimmer and softer as time goes because of the relativistic Doppler 
effect, which explains the observed power-law like decay during 
between 200 and 400~msec very well. 
If the emission is spherical (isotropic), 
however, such a decay should continue beyond $t \gtrsim 600$ msec. This 
is inconsistent with the observation that the light curve rapidly fades 
after $t \sim 600$ msec, which implies that the emission does not occur 
at larger angle from the LOS. In other words, the giant flare arises from 
a relativistic jet with a finite opening angle. 

In order to see the above arguments quantitatively,
we consider a simple model for  emission from a relativistically moving
jet which radiates photons when the shell is located
at radius from $r_0$ to $r_e$ \citep{yamazaki03}.
We assume that the cooling timescale is much shorter than other
timescales, and hence consider an instantaneous emission at the shock front.
We use a spherical coordinate system $(t,r,\theta,\phi)$
in the Lab frame, where the $\theta=0$ axis points 
toward the detector at $r=D$, and the magnetar is located at $r=0$.
The jet has an opening half-angle $\Delta\theta$ and
a viewing angle $\theta_v$  the axis of the emission
cone makes with $\theta=0$ axis.
The emitting shock front moves radially from $t=t_0$ and $r=r_0$
with the Lorentz factor $\gamma=1/\sqrt{1-\beta^2}$.
Then the observed flux per unit frequency 
of a single pulse at the observed time $T$
is given by
\begin{eqnarray}
&& F_{\nu}(T)={{2 {r_0}^{2} \gamma^2}\over{\beta D^2 (r_0/c\beta)}}
\nonumber\\
&& \ \ \times \int dt \, A(t)
{{1-\beta \cos\theta(T)}\over
{1-\beta \cos\theta(t)}}
{{\Delta \phi(t) f[\nu \gamma(1-\beta \cos \theta(t))]}\over
{[\gamma^2(1-\beta \cos\theta(t))]^2}}~~,\nonumber\\
&&
\label{eq:jetthin}
\end{eqnarray}
where $f(\nu')$ represents the spectral shape, and
$1-\beta\cos\theta(T)=(c\beta/r_0)(T-T_0)$,
$1-\beta\cos\theta(t)=[1-\beta\cos\theta(T)]/
[(c\beta/r_0)(t-T_0)]$ and $T_0=t_0-r_0/c\beta$.
The value of $\Delta \phi(t)$ is $\pi$ in the case of
$\theta_v<\Delta\theta$ and $0<\theta(t)\leq \Delta\theta-\theta_v$,
while for the other case, it is given by
\begin{equation}
\Delta\phi(t)=
\cos^{-1}\left[
\frac{\cos \Delta \theta - \cos \theta(t) \cos \theta_v}
{\sin \theta_v \sin \theta(t)}\right]~~.
\end{equation}
The normalization of emissivity $A(t)$ 
is determined by the hydrodynamics.
Here for simplicity we adopt the following functional form,
\begin{equation}
A(t)=A_0\left(\frac{t-T_0}{r_0/c\beta}\right)^{-2}
H(t-t_0) H(t_e-t)~~,
\label{eq:At}
\end{equation}
where the emission ends at $t=t_e$ and
the released energy at each distance $r$ is constant.
Shell emits at radius from $r_0$ to $r_e=\kappa r_0$,
where $\kappa=t_e/t_0>1$.
The quantity $A_0$ is a normalization constant.
A pulse-starting time and ending time are given as
\begin{eqnarray}
&&T_{start}=T_0+({r_0}/{c\beta})
(1-\beta\cos(\max[0,\theta_v-\Delta\theta]))~~,
\label{eq:tstart:xrf}
\\
&&T_{end}=T_0+[({r_0}/{c\beta})+t_e-t_0]
(1-\beta\cos(\theta_v+\Delta\theta))~~,
\label{eq:tend:xrf}
\end{eqnarray}
respectively.
We adopt the following form of the comoving-frame energy spectrum,
\begin{eqnarray}
f(\nu')=(\nu'/\nu'_0)^{1+\alpha}
\exp[-(\nu'/\nu'_0)]~~,
\label{eq:spectrum}
\end{eqnarray}
where $\alpha$ is a power law index.
\citet{mazets05} derived a nonthermal spectrum 
with $\alpha=-0.7$ and an exponential cut-off at 800~keV,
while \citet{palmer05} gave a nonthermal fit with $\alpha=-0.2$ 
and a cut-off at 480~keV.
Although \citet{hurley05} reported a black-body spectrum ($\alpha=1$)
with a temperature higher than $\sim240$~keV,
at least a portion of the giant flare may be nonthermal
because the spectrum was determined with low time resolution.
A previous giant flare in SGR 0526-66 may also have had
a nonthermal spectrum \citep{fenimore96}.
In addition, some giant flares from SGRs in nearby
galaxies (within 40~Mpc) may have been detected by BATSE
as  short GRBs whose spectra are likely
nonthermal \citep{palmer05}.
In this Letter, we assume a nonthermal spectrum, which 
is naturally produced by shocks in a relativistic outflow.

Equations (\ref{eq:jetthin}), (\ref{eq:At}) and (\ref{eq:spectrum})
are the basic equations to calculate the flux of a single pulse,
which depends on following parameters:
$\gamma\theta_v$, $\gamma\Delta \theta$,
$\gamma \nu_0'$, $r_0/c \beta \gamma^2$, $\alpha$, $t_0$, $t_e$, and
$r_0^2\gamma^2A_0/D^2\beta$.
We fix $\gamma\theta_v=0$, $\gamma h\nu'_0=800$~keV,
and $\alpha=-0.6$ in the following.
Then we find out best fit values for other parameters:
$\gamma\Delta\theta$, $r_0/c \beta \gamma^2$, and 
$\kappa~(=t_e/t_0=r_e/r_0)$.
We find that these parameters do not depend on $\alpha$ and $\gamma\nu'_0$ so much,
even if we assume a black-body spectrum.

Figure~\ref{fig:lc} shows the result. 
The fit is surprisingly good considering the very simple model. 
The second pulse, which has a duration of $T_{AB} \sim 50$~msec,
and the associated power-law like decay lasting $T_{BC} \sim 500$~msec,
are well fitted by our model with $\gamma\Delta \theta=3.0$, 
$r_0/\gamma^2=2.6 \times 10^8$~cm, and $\kappa=12.5$.
We have checked that the uncertainty coming from
the spectral shape is at most a factor of 2. 
We also show an example of the theoretical modeling for the first 
pulse ($t<80$ msec) 
by black dotted line in the same figure. 
In this case the opening half-angle is not well 
constrained because the power-law like decay is masked by that of the 
second pulse. 
We can make a similar modeling for a bump at 
$t \sim 430$~msec, though it is not shown in the figure.

The opening angle of the jet is constrained by the light curve in a 
kinematical fashion (see Figures~\ref{fig:lc}, \ref{fig:geo}). The duration 
of the brightest epoch ($80 \lesssim t \lesssim 130$ msec) is 
determined by the crossing time of the shell through the emitting 
region ($r_0<r<r_e$) as 
\begin{equation}
T_{AB} = \frac{(r_e-r_0)(1-\beta)}{c\beta} \sim 
(\kappa-1) \frac{r_0}{2c\gamma^2} \sim 50 {\rm \ msec}.
\end{equation}
On the other hand, the following power-law like decay lasts for 
$\sim 500$ msec, which is approximately given by the angular spreading time
\begin{equation}
T_{BC} = \frac{r_e (1-\cos\Delta \theta)}{c} \sim (\gamma \Delta \theta)^2  
\kappa \frac{r_0}{2c\gamma^2} \sim 500 {\rm \ msec}.
\end{equation}
In other words the wider the jet, the later the onset of the steep decay. 
Eliminating $r_0/2c\gamma^2$ from these equations, we derive 
$(\gamma \Delta \theta)^2 \sim 10(1-\kappa^{-1}) \lesssim 10$ and 
hence $\gamma \Delta \theta \lesssim 3$. The uncertainty is at most a factor 
of $2$. Furthermore, combining with $\gamma \gtrsim 25$ required to avoid 
pair formation \citep{nakar05}, we obtain a firm upper limit, 
$\Delta \theta \lesssim 0.2$ radian, which is very similar to those of 
GRB jets inferred from the observed break in afterglow light curves
\citep{harrison99}. In particular, we disfavor models with  isotropic 
 emission for the beginning epoch of the giant flare. 

\begin{figure}[t]
\begin{center}
\FigureFile(80mm,30mm){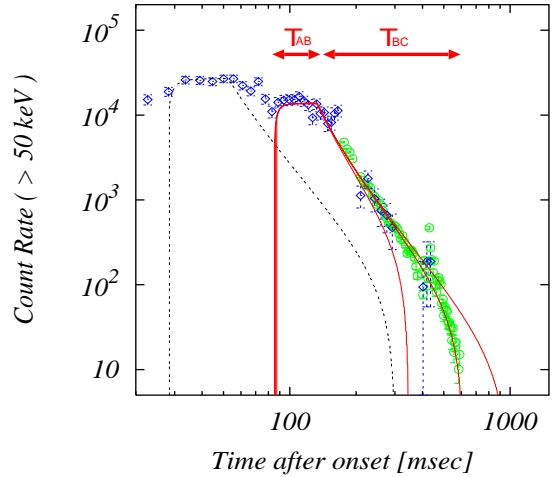}
\end{center}
\caption{
Comparison of theoretically predicted light curves with the observed 
data. Green and blue points are background-subtracted
MCP and CEM data, respectively \citep{terasawa05}. 
The second pulse and the following power-law like 
decay are modeled by red lines, which have 
$\Delta \theta =2 \gamma^{-1}, 3 \gamma^{-1}$, and $4 \gamma^{-1}$ 
from left to right with $r_0=2.6 \times 10^8 \gamma^2$ cm and 
$r_e=12.5r_0$. The rapid fading at $t\sim 600$ msec is most consistent 
with an opening half-angle of $\Delta \theta=3 \gamma^{-1}$. 
The black dotted line shows
an example of the theoretical modeling for the first pulse ($t<80$ msec).}
\label{fig:lc}
\end{figure}

\begin{figure}[t]
\begin{center}
\FigureFile(80mm,50mm){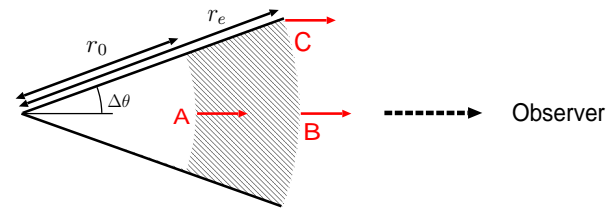}
\end{center}
\caption{A schematic picture of the jet emission.
An observer resides 
far on the right side. A thin shell emits gamma-rays while it crosses 
the hatched region ($r_0<r<r_e$). Each red arrow represents the emitted 
photon at each place. The observed duration $\sim 50$ msec of the second pulse 
($80<t<130$ msec) in Figure~\ref{fig:lc} is determined by the shell 
crossing time, $T_{AB}$, i.e., the difference of the arrival time of 
two photons A and B that are emitted when the shell crosses radii 
$r_0$ and $r_e$, respectively. The observed duration of the power-law 
like decay after the second pulse ($130<t<600$ msec) in Figure~\ref{fig:lc} 
is determined by the angular spreading time, $T_{BC}$, i.e., the difference 
of the arrival time of two photons B and C emitted simultaneously. 
The wider the jet is, the later the rapid fading begins.}
\label{fig:geo}
\end{figure}

\section{Discussions}

We have shown that the initial spike of the giant flare SGR~1806-20
is well explained by emission from a relativistic jet directed toward us,
and that the opening half-angle of the jet is less than 0.2~rad.
Since the isotropic equivalent energy of the giant flare is 
$5 \times 10^{46}$ ergs with assumed distance of 15~kpc \citep{terasawa05}, 
the collimation-corrected energy is less 
than $5 \times 10^{44}$ ergs for $\Delta \theta < 0.2$ and the flare 
is rather economical than previously thought. This may alleviate an 
extreme situation that an isotropic flare demands almost all energy 
of dipole magnetic fields of SGR~1806-20. 
The size and the light curve of 
the radio afterglow from SGR~1806-20 also favor smaller energy 
$\sim 10^{44}$-$10^{45}$ ergs \citep{cameron05,gaensler05}
than the isotropic equivalent energy of the flare,
which also suggests a jet opening half-angle 
$\Delta \theta \sim 0.2$. 

External shock scenario is unlikely
because it is  difficult  to keep high Lorentz factor and have a
steep decay at 500~msec.
The third bump at $t\sim 430$ msec may be caused by an 
additional internal shock, while other reasons such as inhomogeneities 
on the jet are possible since its peak flux is much smaller than the others.

A relativistic jet begins sideway expansion when the jet 
 Lorentz factor becomes $\gamma\sim(\Delta\theta)^{-1}$. 
This epoch is observed at $T_{jet}\sim9~{\rm min}~(E_{45}/n_0)^{1/3}
(\Delta\theta/0.1~{\rm rad})^2$, where 
$E_{45}$ and $n_0$ is the total energy confined in the jet and
the number density of the ambient matter.
After that time, the jet decelerates abruptly.
Hence in the epoch of radio observations (6--20~days after the flare),
the outflow became Newtonian or sub-relativistic and nearly isotropic,
which is consistent with the radio observation.
Nevertheless some degree of anisotropy may remain and
produce the observed elliptical image and polarization of the radio afterglow 
\citep{gaensler05}.

It has been discussed that giant flares from other SGRs resemble 
classical GRBs in spectroscopic characters \citep{fenimore96}. This 
possibility is strengthened by our present result that the recent 
giant flare of SGR~1806-20 is a jetted emission like GRBs. However, in our
scenario, there should be many more misaligned SGRs, which will show 
up only in isotropic emission. If the pulsating tails are isotropic 
emission, there should be many events consisting of only pulsating 
tails, though such events that are bright enough to trigger
e.g. BATSE have not yet been reported.
One possibility to  resolve this ``statistical'' problem
is that the pulsating tail is also collimated.
According to the magnetar model, pulsating tail emission arises
from an evaporating trapped fireball \citep{thompson95,thompson01}.
Its size can be comparable to the radius of the magnetar.
However, as discussed in \citet{thompson01}, due to the QED effect,
the radiative transport across the magnetic field lines is concentrated
to the foot-point at the magnetar surface, which may be a reason for
the collimation of the pulsating tail.
The collimation angle would largely depend on the magnetic field
configuration.
However, the broad pulse profile (larger than 1~sec) of the tail disfavors
the narrowly collimated pulsating tail emission. 
Another way to resolve the problem is to introduce some envelope
around the main jet.
In the GRB case, it is now widely argued whether the angular structure
of the jet is uniform, Gaussian, power-law, or two-component 
\citep{zhang04}.
Similar to the GRB, it may be possible that the SGR jet
also has a central core with $\Delta\theta\sim0.1$ and an envelope with wider
solid angle that can produce less energetic flares.
Indeed, the past two giant flares show much smaller isotropic equivalent
energies, $\sim10^{44}$~ergs, while
all three flares had pulsating tails with a similar energy $\sim10^{44}$~ergs.
If the central core is seen off-axis and the observer points to the 
envelope, such less energetic giant flares can be observed.
At that time, the intense emission from the central core may be negligible
because of the relativistic beaming effect.
If such an envelope has a wide solid angle,
the statistical problem can be resolved.
Indeed, exponentially decaying tail after $t\sim600$~msec recorded by 
BAT/{\it Swift} \citep{palmer05} may be such an envelope emission.

The relativistic jet may not be generally launched along 
the rotation axis because the magnetic energy dominates over
the rotational energy.
The duration for the magnetar to eject a jet 
is less than the duration of a pulse, $t_{dur}<T_{AB}\sim50$~msec.
Thus the rotation angle during the jet ejection is at most
$\vartheta_{rot}\sim2\pi(t_{dur}/T_p)\lesssim0.04(t_{dur}/50~{\rm msec})$~rad,
where $T_p=7.56$~sec is the rotational period of SGR~1806-20
\citep{hurley05}.
Therefore, the overall opening half-angle is at most
$\Delta\theta+\vartheta_{rot}/2$.
Since this is comparable to $\Delta\theta$,
the effect of the rotation of the magnetar may not be so large
but may still affect the jet structure and generate the envelope
as discussed in the previous paragraph.

If the energy source of the flare was a dipole outer magnetic field
and $\sim 10\%$ of its energy was converted to $\gamma$-rays
as in the case of the isotropic flare,
the spin down rate, $\dot{P}\propto B^2/P$, where $P$ is the rotational
period of the magnetar,  would change  by $\sim 10\%$ after the flare.
Thus the isotropic flare might be tested by the 
observations of the spin down history.

It is widely believed that anomalous X-ray pulsars (AXPs) are the same
kinds of objects as SGRs (e.g., \cite{gavriil04}).
\citet{kulkarni03} proposed the main difference between AXPs and SGRs
is the time-dependent geometry of the magnetic fields, i.e.,
AXPs are older and less active than SGRs.
Here we suggest that the viewing angle may also contribute to
the different appearances of AXPs and SGRs.
If we see the passive (active) part of the magnetar surface,
the magnetar would show soft (hard) persistent X-ray emissions
and look like AXP (SGR).
This might be consistent with no giant flares from AXPs 
because the energetic jet core responsible for the giant flare is 
always misaligned.

\bigskip

We thank T.~Terasawa for useful comments and providing
GEOTAIL data.
This work was supported in part by 
JSPS Research Fellowship for Young Scientists (RY),
the Eberly Research Funds of Penn State 
and the Center for Gravitational Wave Physics under grants 
PHY-01-14375 (KI).



\begin{thebibliography}{}
%
\bibitem[Boggs et al.(2004)]{boggs04}
Boggs,~S. et al. 2005,  GRB Coordinates Network, 2936
%
%
\bibitem[Cameron et al.(2005)]{cameron05}
Cameron,~P.~B. et al. 2005, astro-ph/0502428
%
\bibitem[Cline et al.(1980)]{cline80}
Cline, T. et al. 1980, ApJ, 237, L1
%
\bibitem[Fenimore, Klebesadel \& Laros(1996)]{fenimore96}
Fenimore,~E.~E., Klebesadel,~R.~W., \& Laros,~J.~G. 1996, ApJ 460, 964
%
\bibitem[Frail, Kulkarni \& Bloom(1999)]{frail99}
Frail,~D., Kulkarni,~S.~R. \& Bloom,~J. 1999, Nature, 398, 127
%
\bibitem[Gaensler et al.(2005)]{gaensler05}
Gaensler, B. M. et al. 2005, astro-ph/0502393
%
\bibitem[Gavriil et al.(2004)]{gavriil04}
Gavriil, F. P., Kaspi, V. M., \& Woods, P. M. 2004, 
Adv. Sp. Res. 33, 654
%
\bibitem[Harrison et al.(1999)]{harrison99}
Harrison,~F.~A. et al. 1999, ApJ, 523, L121
%
\bibitem[Hjorth et al.(2003)]{hjorth03}
Hjorth,~J. et al. 2003, Nature, 423, 847
%
\bibitem[Huang, Dai \& Lu(1998)]{huang98}
Huang,~Y.~F., Dai,~Z.~G. \& Lu,~T. 1998,
Chinese Physics Letters 15, 775
%
\bibitem[Hurley et al.(1999)]{hurley99}
Hurley, K. et al. 1999, Nature, 397, 41
%
\bibitem[Hurley et al.(2005)]{hurley05}
Hurley, K. et al. 2005, astro-ph/0502329
%
\bibitem[Kulkarni et al.(2003)]{kulkarni03}
Kulkarni, S. R. et al. 2003, ApJ, 585, 948

\bibitem[Mazets et al.(2005)]{mazets05}
Mazets, E. P. et al. 2005, astro-ph/0502541
%
\bibitem[Nakar, Piran \& Sari(2005)]{nakar05}
Nakar,~E., Piran,~T. \& Sari,~R. 2005, astro-ph/0502052
%
\bibitem[Palmer et al.(2005)]{palmer05}
Palmer, D. M. et al. 2005, astro-ph/0503030
%
\bibitem[Piran(1999)]{piran99}
Piran,~T. 1999, Phys.~Rep., 314, 575
%
\bibitem[Piran(2004)]{piran04}
Piran,~T. 2004, Rev.~Mod.~Phys., 76, 1143
%
%
\bibitem[Stanek et al.(2003)]{stanek03}
Stanek,~K.~Z. et al. 2003, ApJ, 591, L17
%
\bibitem[Terasawa et al.(2005)]{terasawa05}
Terasawa,~T. et al. 2005, astro-ph/0502315
%
\bibitem[Thompson \& Duncan(1995)]{thompson95}
Thompson,~C. \& Duncan,~R.~C. 1995, MNRAS, 275, 255
%
\bibitem[Thompson \& Duncan(2001)]{thompson01}
Thompson,~C. \& Duncan,~R.~C. 2001, ApJ, 561, 980
%
\bibitem[Wang et al.(2005)]{wang05}
Wang,~X.~Y. et al. 2005, ApJL in press (astro-ph/0502085)
%
\bibitem[Woods(2004)]{woods04}
Woods, P. M. 2004, Adv. Sp. Res. 33, 630
%
\bibitem[Yamazaki, Ioka \& Nakamura(2003)]{yamazaki03}
Yamazaki,~R., Ioka,~K. \& Nakamura,~T. 2003, ApJ, 591, 283
%
\bibitem[Zhang \& M\'{e}sz\'{a}ros(2004)]{zhang04}
Zhang,~B. \& M\'{e}sz\'{a}ros,~P. 2004,
Int.~J.~Mod.~Phys. A19, 2385
%
\end{thebibliography}
\end{document}